\documentstyle[psfig,amsmath,eqsecnum,multicol,aps,prb]{revtex}

\title{Diffusion as  mixing mechanism in granular materials}
\author{C. Henrique$^1$, G. Batrouni$^2$ and D. Bideau$^1$}
\address{$^1$ Groupe Mati\`ere Condens\'ee et Mat\'eriaux (UMR
6626), Universit\'e de Rennes I, B\^at 11A, Campus de Beaulieu 35042
Rennes Cedex, France}
\address{$^2$ Institut Non-Lin\'eaire de Nice, Universit\'e de Nice-Sophia
Antipolis, 1361 route des Lucioles, 06560 Valbonne, France}

\begin{document}
\maketitle
\begin{abstract}
We present several numerical results on granular mixtures. In
particular, we examine the efficiency of diffusion as a mixing
mechanism in these systems. The collisions are inelastic and to
compensate the energy loss, we thermalize the grains by adding a
random force. Starting with a segregated system, we show that uniform
agitation (heating) leads to a uniform mixture of grains of different
sizes. We define a characteristic mixing time, $\tau_{mix}$, and study
theoretically and numerically its dependence on other parameters like
the density.  We examine a model for bidisperse systems for which we
can calculate some physical quantities. We also examine the effect of
a temperature gradient and demonstrate the appearance of an expected
segregation.
\end{abstract}
\section{Introduction}
\label{sec:intro}
Granular media are notoriously difficult to mix. For a variety of
reasons and under rather general conditions, they tend to form
segregated steady states~\cite{guyon}. For example, segregation can
occur during granular flow, the bigger particles moving farther than
the smaller ones. Difference of size (polydispersity of the grains) or
difference of material (different kinds of grains) produce different
geometrical or physical properties.  Segregation can also be due to
percolation~\cite{lili} where the small grains fall through the holes
between the big grains leaving only the bigger particles behind.
Shear~\cite{scott,campbell,savage} and vibration can also produce
segregation. One of the best known examples of vibration segregation
is perhaps the ``Brazil nut effect''. In this case, the geometrical
properties~\cite{rosato,julien,duran} are responsable for the upward
movement of big particles although convection processes near the
boundaries can also be very important~\cite{knight}. All these
processes (flow, shearing and convection) are very common in
industrial applications such as in mixers~\cite{lacey,fan}. For these
reasons such mixers are efficient only for rather homogeneous
materials. In the polydisperse case it is very hard to avoid
segregation.

For gases and liquids, the thermal agitation of molecules is a natural
and efficient mechanism leading to throughly mixed systems with
homogenous equilibrium steady states~\cite{hulin}. We propose here, a
study of a system of agitated grains in analogy with liquid or gas
molecules at the miroscopic scale.

Two major differences between granular materials and fluids are: (1)
the particle size compared to the mean free path and (2) the inelastic
properties and friction responsable for energy dissipation during
collisions. The question then is if these differences will alter the
system's natural tendency to mix by diffusion. In other words, is it
possible to use diffusion to mix grains. In spite of the dissipative
collisions, it is possible to keep a granular system agitated, for
example on an air-table or a vibrating bed.  To simulate numerically
such constantly agitated granular systems, we add an external random
force to the equations of motion (see section~\ref{sec:num}). We then
analyze the grain diffusion and its dependence on the various
parameters of the system such as grain size. We find that inspite of
the dissipative nature of the collisions, diffusion is still a good
mixing mechanism just like in fluids.

In section~\ref{sec:num} we detail the algorithm and summarize the
principal dynamic equations and parameters of our system. We verify
our procedure with the study of a monodisperse system in
section~\ref{sec:mono} and etablish relations which characterize the
temporal evolution of an initially segregated system. The bidisperse
case is studied in section~\ref{sec:bidi}. We show in particular the
evolution of a system with homogeneous agitation, and also the effects
of a gradient in this agitation. Our conclusions and discussion are in
section~\ref{sec:conclu}.

\section{Algorithm and review}
\label{sec:num}
We use an Event Driven Molecular Dynamics algorithm. The simulated
grains have the same characteristics (e.g. normal and tangential
restitution coefficients) as measured experimentally~\cite{loudge}.
To thermalize the system, we add at regular time step intervals, $dt$,
external random forces which act on every particle. There are several
choices one can make for this force. Our choice is the following:
\begin{equation}
\label{eq:f^t}
F_i^t= m \left(\sqrt{\eta_0^2/dt}\right)\zeta_i
\end{equation}
with $i=x,y$ (corresponding to the two directions), $\zeta_i$ is a
gaussian noise caracterized by $<\zeta_i\zeta_j>=\delta_{i,j}$ and $m$
is the particle mass. $\eta_0^2$ is the control parameter which we use
to increase or decrease the agitation.  Experimentally, on the air
table\cite{jacques}, plastic disks of radius $R$ move due to the
fluctuations of the air flux acting on their surfaces. Therefore,
since their mass is proportional to $R^2$, we expect the acceleration
to be independent of $R$. That is why we have chosen an external force
proportional to the particle mass, Eq.~(\ref{eq:f^t}). The equation of
motion of a particle between two collisions becomes:
\begin{equation}
\label{eq:newton}
\frac{d \mathrm{v_{i}}}{dt}= \sqrt{\eta_0^2}\zeta_i. 
\end{equation}
In this paper $\mathrm{v_i}$ denotes the instantaneous velocity in the
$\mathrm{i}$ direction and $v^2$ the mean square velocity.  The system
is two dimensional and is enclosed in a square box whose walls are
made of grains of radius $r_w$ and are infinitely massive. The
particle-wall collisions are taken to be elastic. Note that in these
two-dimensional simulations, the particles are represented as spheres
interacting at their equators.

\subsection{Macroscopic characteristics of the steady state}
\label{sec:thermi}

The above model leads to a steady state characterized by a constant
mean square velocity, $v^2(t\to \infty)$ for all particles. The energy
loss during collisions is compensated for by the random force.  For a
monodisperse gas, the energy balance is easily calculated. The energy
loss, $\Gamma$, per unit time in the steady state is given
by~\cite{trizac}

\begin{equation}
\Gamma \propto  \omega m v^2,
\end{equation}
where $\omega$ is the frequency of collisions.  On the other hand, the
average gain in energy due to the random force during $dt$ is easily
obtained from Eq.~(\ref{eq:newton}):
\begin{equation}
\frac{1}{2}m\left[v^2(t+dt) -v^2(t)\right] = m\eta_0^2dt.
\end{equation}
In the steady state we can thus write
\begin{equation}
\label{eq:balance}
\frac{1}{2}m\frac{\partial v^2}{\partial t} = - \Gamma + m\eta_0^2.
\end{equation}
We can write $\omega \sim \tfrac{\sqrt{v^2}}{l}$ where $l$ is a
caracteristic length depending only on the packing fraction and on the
radius of the grains.  With this assumption and the fact that
$\frac{\partial v^2(t \to \infty)}{\partial t}=0$,
Eq.~(\ref{eq:balance}) gives
\begin{equation}
\label{eq:v2-l-eta}
v^2(\infty) \propto (l\eta_0^2)^{2/3}.
\end{equation}
This power law is independent of the coefficients of restitution and
friction if dissipation is not too large. The results of the kinetic
theory of inelastic gases can be applied, in particular the velocity
distribution can be approximated by a Maxwellian.  The parameter
$\eta_0^2$ allows us to change the granular temperature, $T$, since
$v^2 \propto T$. Therefore, in the steady state, $T$ is independent of
the initial conditions. For polydisperse gases the problem becomes
more complicated as will be seen below.

\subsection{Coefficient of diffusion}
\label{sec:diff}
Since the mean square velocity is constant, and consequently the
collision frequencies too, particles have a simple diffusive behavior.
The mean square displacement, $<(r(t+t_0) -r(t_0))^2>$, gives the
coefficient of diffusion
\begin{equation}
\label{eq:defD}
   <(r(t+t_0) -r(t_0))^2> = 4Dt,
\end{equation}
where $t_0$ is large enough to ensure thermalization of the system.
Clearly, if the system is examined at a time scale $t \le 1/\omega$ we
will not observe the true diffusive behaviour.  At short time,
$\mathrm{v_i}$ is not constant due to the action of $F_i^t$ and so the
mean square displacement is not yet linear with $t$.  For $t >>
1/\omega$ we verify the linear dependence of the mean square
displacement on time, for all particles. The value of the coefficient
of diffusion $D$ found from the simulation is thus larger than the
theoretical value predicted by a Langevin description due to the
dynamics of the particle at short time~\cite{boundaries}.

\subsection{Simulation procedure}
\label{sec:mixt}
We study systems made of two species of grains, $s$ and $b$. The
radius of the particles are respectively $R_s$ and $R_b$.  The system
is a square box of length $L$ and we use the boundary conditions
discussed above. The number of particles of each species is calculated
based on the desired packing fraction $C$ and the relative proportion
of $s$ particles $x_s$.
\begin{equation}
\label{eq:def-xs-C}
\left\lbrace
\begin{array}{l}
C=\frac{n_s \pi R_s^2 + n_b \pi R_b^2}{L^2},\\
x_s= \frac{n_s \pi R_s^2}{ n_s \pi R_s^2 + n_b \pi R_b^2},
\end{array}
\right.
\end{equation}
where $n_s$ and $n_b$ are repectively the number of particles $s$ and
$b$.

For all mixtures, we performed two types of simulations. In the first
one, the two species $s$ and $b$ are already mixed and the initial
position of each particle is chosen randomly in the box by using a
classical algorithm of Random Sequential Adsorption (R. S. A.)
\cite{rsa}.  We are careful that this algorithm does not introduce
segregation in the initial configuration.  For the second type of
simulation, the two species are initially separated with the $s$
particles on the right and the $b$ particles on the left. The system
is prepared such that the packing fraction is homogeneous in the whole
system.

In the following section we present our results in the simple case
where the two species have the same mechanical and geometrical
properties, i.e. $s$ and $b$ grains are of the same type.

\section{Monodisperse case}
\label{sec:mono}

To test the validity of our algorithm we start with identical grains,
i.e. $R_s=R_b$. In this case, we do not expect segregation because the
grains are identical, but we would like to verify that the thermal
process is efficient and that particles do not agregate. In other
words, after some time each grain will have visited all regions of the
box. In figure \ref{fig:mono} is presented the temporal evolution of
the system for the two different initial configurations specified
above.  In figure \ref{fig:mono}{\bf a} the system is already mixed
and in \ref{fig:mono}{\bf b} the particles are initially separated.
As one can see in this figure, the system does not collapse and the
grains are homogeneously distributed in the box. To analyse the
dynamics of the mixture we measure the quantity $N_{s,b}(t)$ defined
as the number of collisions between $s$ and $b$ grains per unit time.
The evolution of $N_{s,b}(t)$ with time gives two important results.
For the initial configuration corresponding to figure
\ref{fig:mono}{\bf a}, the quantity $N_{s,b}(t)$ fluctuates around a
mean value, $N_{s,b}(\infty)$, as seen in figure
\ref{fig:N-t_mono}. This means that the system has reached a steady
state in which the mean square velocity of the particles is constant
and equal to $v^2(\infty)$. Note that the system evolves very
quickly into this steady state. We have checked that all the
configurations at different times, $t$, are statistically identical
and that the system remains homogeneous. There is no evidence of
collapse or cluster formation. The second observation is that for the
initial configuration corresponding to figure \ref{fig:mono}{\bf b},
the quantity $N_{s,b}(t)$ increases and then stabilizes at large time
at the value $N_{s,b}(\infty)$ defined above. However, the mean square
velocity of the grains, $v^2(t)$, reaches the steady state value
$v^2(\infty)$ much more quickly since the grains are identical. The
knowledge of the mean square velocity is not sufficient to define the
state of the system since it gives no information about the spatial
repartition of the two species.  $N_{s,b}(t)$ is therefore the only
pertinent quantity to characterize the homogeneity of the system.

A few more comments about $N_{s,b}(t)$ are in order.  In the system
studied above, the packing fraction and the velocity distribution are
spatially homogeneous and constant in time (except at very short
time). As a consequence the quantity $N_{s,b}(t)$ depends only on the
spatial repartition of the two types of grains. The evolution of
$N_{s,b}(t)$ allows us to define a mixing time, $\tau_{mix}$.  We have
already mentioned that the velocity is the same for all particles and
independent of position. This is also true for the local density.  We
conclude that also the frequency of collisions is the same for all
particles and is space independent.  In this monodisperse case, the
dynamics are purely diffusive and can be characterized by a
coefficient of diffusion $D$ which is independent of horizontal
spatial position $x$. Let us call $\delta N_{s,b}(x,t)$ the number of
collisions between the two species occuring at a position between $x$
and $x+dx$ at time $t$.  Clearly, $\delta N_{s,b}(x,t)$ is directly
proportionnal to $d_s(x,t)$ and $d_b(x,t)$, the densities of $s$ and
$b$ grains at position $x$. The densities $d_s$ and $d_b$ do not
depend on the vertical position since the system is invariant along
this direction. We will define $d_0$ as the total density and can,
thus, write $d_0 = d_s(x,t) + d_b(x,t)$. $d_0$ is of course
independent of $x$ because the system remains homogeneous.  We then
obtain an expression for $N_{s,b}(t)$:
\begin{equation}
\label{eq:N-t}
N_{s,b}(t) \propto \int\limits_{0}^{L}d_b(x,t)(d_0-d_b(x,t))dx.
\end{equation}
The density of big particles at $(x,t)$, $d_b(x,t)$, is described by
Fick's equation,
\begin{equation}
\label{eq:fick}
\frac{\partial d_b(x,t)}{\partial t}= D 
\frac{\partial^2 d_b(x,t)}{\partial x^2},
\end{equation}
with the following boundary conditions:
\begin{equation}
\label{eq:limits}
\left\lbrace
\begin{array}{ll}
d_b(x)=d_0 \quad & \text{for $0 \le x < L/ 2$ and $t=0$}, \\
d_b(x)=0 \quad  & \text{for $L/2 \le x  \le L$  and $t=0$}, \\
d_b(x)=d_0/2 \quad & \text{for all $ x$  and $t\simeq \infty$}.\\
\end{array}
\right.
\end{equation} 
We assume the solution of Eq.~(\ref{eq:fick}) has the form:
\begin{equation}
\label{eq:soldb}
d_b(x,t)=\sum\limits_{m=0}^{\infty}(B_msin(\lambda_mx) +
A_mcos(\lambda_mx))exp(-\lambda_m^2Dt) + \frac{d_0}{2}
\end{equation}
where the $\lambda_m$ are constants. Using the conditions
Eq.~(\ref{eq:limits}) gives:
\begin{equation}
\label{eq:soldb2}
d_b(x,t) =
\sum\limits_{k=0}^{\infty}a_kcos(\lambda_kx)exp(-\lambda_k^2Dt) +
\frac{d_0}{2}
\end{equation}
with 
\begin{equation}
\label{eq:lambdaK}
\begin{array}{l}
a_k=\frac{2d_0(-1)^{-k}}{\pi(2k+1)},\\
\lambda_k= \frac{(2k+1)\pi}{L}.
\end{array}
\end{equation}
Eq.~(\ref{eq:N-t}) then gives the final
expression for $N_{s,b}(t)$,
\begin{equation}
\label{eq:N-solv}
\begin{array}{c}
N_{s,b}(t) \propto \frac{d_0^2L}{4}\left(1 -\sum\limits_{k=0}^{\infty}
\frac{8}{\pi^2(2k +1)^2}exp(-2 \lambda^2_kDt)\right),\\
\lambda_k^2=\frac{(2k+1)^2\pi^2}{L^2}.
\end{array}
\end{equation}
As a first approximation, we may keep only the first mode and write:
\begin{equation}
 \label{eq:N-exp}
\begin{array}{c}
N_{s,b}(t)\simeq N_{s,b}(\infty)\left(1 -exp(-t/\tau_{mix})\right), \\
\tau_{mix}= \frac{L^2}{2\pi^2D},
\end{array}
\end{equation}
where $\tau_{mix}$ can be taken as the typical time for mixing.  

To check the validity of the theoretical expressions for $N_{s,b}(t)$
and $\tau_{mix}$ established above, we have performed simulations for
different values of $x_s$ and $C$. For a given set of parameters, we
have performed five simulations corresponding to different initial
positions and velocities of the particles for the case where the $s$
and $b$ grains are initially separated. Figure \ref{fig:n-fit} shows
$N_{s,b}(t)$ versus $t$ averaged over the 5 simulations. We see that
the agreement between theory and numerical simulation is very good.
Figure \ref{fig:dep-D} shows the dependence of the mixing time,
$\tau_{mix}$, on the coefficient of diffusion $D$. To get this, we
performed several simulations changing the packing fraction and the
radius of the particles in order to vary the coefficient of diffusion.
Note that $D$ was estimated using Eq.~(\ref{eq:defD}). The slope of
the curve is exactly that predicted by the theory.  As the coefficient
$D$ can be calculated from the parameters of the system ($R_s=R_b$,
$\eta_0^2$, $C$), we can estimate analytically and with high accurancy
the mixing time. We have also verified the dependence of $\tau_{mix}$
on $L^2$, Eq.~(\ref{eq:N-exp}), and have found very good agreement
too.

\section{Bidisperse case}
\label{sec:bidi}

We now discuss the case of a binary mixture. We will see that the
size difference between the grains changes drastically the dynamics of
the system.  The grains $s$ and $b$ are taken of equal density and
identical coefficients of restitution and friction, and we take $R_s <
R_b$.

We present first the case where the system is thermalyzed uniformly,
{\it i.e.} $F_i^t$ does not depend of the position of the grain. Then
we will examine the case where a gradient is imposed on the agitation
force.

\subsection{Case of homogeneous agitation}
\label{sec:bidi-H}

We will show that, in the bidisperse case, the system also evolves
into a homogeneous steady state. We will see here that the form of the
thermalization force and the initial conditions determine the
evolution of the system towards the steady state. In the simulations the
packing fraction is fixed to $40 \%$ and $x_s$, which represents the
relative proportion of small grains (Eq.~(\ref{eq:def-xs-C})), is
the only parameter to be varied. 

\subsubsection{Evolution at short time}
\label{sec:shortT}

Figure \ref{fig:etero} shows the evolution of the system with time
$t$. In the initial configuration, the two species are separated and
the two populations have the same initial velocity distribution and
therefore $v^2_s=v^2_b$.  The initial local packing fraction, as one
can see in figure \ref{fig:etero}, is the same in the whole box.

Recall that in our simulations the surface occupied by a particle in
the plane is proprotional to $R^2$ and its mass to $R^3$ since the
particles are spherical. Since the pressure is $P \propto m d v^2$,
where $d$ is the density of grains, the initial pressure is larger for
the bigger particles.  The system therefore has an initial pressure
difference which will govern its behaviour immediately after the
partition is removed whereby the bigger particles, $b$, compress the
small ones $s$. As $t$ increases, (see figure
\ref{fig:etero}) the density of the $b$ particles decreases and so does
its pressure. On the other hand, the density and pressure of the $s$
particles increase.  During this compression period we can consider
the system as two interacting monodisperse systems.  In the left part
of the box (occupied by the larger particles), as the density
decreases, the mean free path, $l$ increases. We have seen in section
\ref{sec:thermi} (see eq \ref{eq:v2-l-eta}) that $v^2$ increases with
$l$.  The velocity $v^2_b$ is then increasing with time. For the same
reason, in the right part of the box, $v^2_s$ is decreasing. As a
consequence, the pressure, which is proportional to the product of the
square velocity and the density, is maintained almost constant in each
subsystem. The pressure difference between the two subsystems remains
therefore important and favors the compression of small particles. The
packing fraction of the $s$ grains increases up to a value around $68
\%$.

It is worth noting that if the walls were inelastic, collapse would
occur whereby the small grains would be squeezed near the wall and
would loose all their energy due to dissipation.  In our simulations
we use elastic wall and thus observe a reflection of the compression
wave. To illustrate this we show in figure \ref{fig:vague} the packing
fraction of the $s$ grains as a function of $t$ and $x$.  We observe a
compression wave which traverses the system. On average, the small
particles remain compressed and the big ones dilute. During this
process the diffusion between the big and the small particles is very
efficient due to the high concentration gradient. The evolution of the
quantities $v^2_s(t)$ and $v^2_b(t)$ as a function of time is
illustrated in figure \ref{fig:vit-t}. One can see that the velocity
of small particles decreases at short time and then increases when the
mixing process starts. At long time the mean square velocities of both
species reach a constant value corresponding to a steady state.

\subsubsection{Mixing time}
\label{sec:bidi_tm}

After the compression phase, the system starts to mix.  As we have
done for the monodiperse case we examine the quantity $N_{s,b}(t)$. To
have a good estimate of $N_{s,b}(t)$, we take (as in the previous
section) the mean value obtained over 5 simulations. Figure
\ref{fig:N-t_etero} shows that we can appoximate $N_{s,b}$ by
$N(\infty)exp(-t/\tau_{mix})$.  Note that the compression phase occurs
during a short time compared to $\tau_{mix}$. We obtain in this way
$\tau_{mix}$ for different values of $x_s$ for $R_s=0.4$ and
$R_b=0.6$.  It appears that this time $\tau_{mix}$ can be considered
as independent of $x_s$ (see Figure \ref{fig:t_mixt_etero}).  This
mixing time obtained in a bidisperse system is smaller than that
obtained for the same packing fraction in the monodisperse case.

\subsubsection{Steady state}
\label{sec:largeT}

Even though the collisions are dissipative, the system does reach an
out of equilibrium stationary state due to the random agitation force.
This stationary state should be characterized by macroscopic functions
which should be independent of time.

Using thermalized configurations (long evolution times), we performed
a geometrical analysis using Vorono{\"\i} tessellation to check that
no segregation exists. We calculated the number of $s$ neighbours for
a $b$ particle and found that the distributions of neighbours are
roughly identical to the distributions obtained from static
configurations generated by an R.S.A. algorithm.  We should point out
that the distances between particles can be different from the static
case but the neighbourhood of a grain is the same in the dynamic and
static situations. This demonstrates that there is no segregation.

We now consider the distribution of the kinetic energy as a function
of the radius of grains.  In the case of elastic collisions, we can
define a kinetic temperature $T$ even in a polydisperse case.  In a
forced inelastic system the repartition of energy seems to be very
different and depends on the type of forcing used\cite{ire,yann}.

In our system, the mean square velocity $v_i^2$ of particle $i$
depends on its mass $m_i$ and also on the proportion of all species
$j$ and their masses $m_j$. In all cases $v^2_i$ is constant at large
time. Energy balance in a bidisperse system means that the agitation
energy per unit time for a given species equals the energy lost in
collisions with particles from all species. This can be written as
follows:
\begin{equation}
\label{eq:thermo1}
\left\lbrace
\begin{array}{l}
P(m_s,m_s)w_{ss} m_s v_s^2 + P(m_s,m_b)w_{sb} m_s v_s^2 = m_s\eta_0^2,
\\ P(m_b,m_b)w_{bb} m_b v_b^2 + P(m_b,m_s)w_{bs} m_b v_b^2 = m_b
\eta_0^2,
\end{array}
\right.
\end{equation}
where $P(m_i,m_j)$ is the mean relative loss of kinetic energy by
particle $i$ when particles $i$ and $j$ collide. Clearly, this term
depends both on mass and relative velocity.  We also expect that
$P(m_s,m_s)$ should be equal to $P(m_b,m_b)$.  Therefore the
quantities $P(m_s,m_s)$ and $P(m_b,m_b)$ calculated in a bidisperse
system must be the same as those obtained for the monodisperse
case. This term $P(m_i,m_i)$ is therefore only a function of
coefficients of restitution and friction. The term $w_{ij}$ represents
the frequency of collisions of the grains $i$ with $j$ grains.  We
note that $n_s w_{sb} =n_b w_{bs}$. Using Enskog theory
\cite{chapman}, the frequencies of collisions of the grains in our
two-dimensional system are given by:
\begin{equation}
\label{eq:thermo2}
\left\lbrace
\begin{array}{l}
w_{ss} = \chi \sqrt{\pi}(2R_s) \frac{C x_s}{\pi R_s^2}
\sqrt{2v_s^2},\\ w_{sb} = \chi \sqrt{\pi}(R_s + R_b) \frac{C
(1-x_s)}{\pi R_b^2} \sqrt{v_s^2 + v_b^2},\\ w_{bb} = \chi
\sqrt{\pi}(2R_b) \frac {C (1- x_s)}{\pi R_b^2} \sqrt{2v_b^2}, \\
w_{bs} = \chi\sqrt{\pi}(R_s + R_b)\frac{C x_s}{\pi R_s^2} \sqrt{v_s^2
+ v_b^2}.
\end{array}
\right.
\end{equation}
$\chi$ is a correction factor and corresponds to the local radial
distribution around a particle \cite{jenkin,verlet}. We have
previously shown in \cite{moi-t} that $\chi$ does not depend on the
type of particles but only on the packing fraction.

The two limit cases, $x_s=0$ and $x_s=1$, correspond to monodisperse
situations with $R=R_s$ and $R=R_b$ respectively.  In these two cases,
we have determined numerically the four parameters $v^2_s(x_s=0)$,
$v^2_s(x_s=1)$, $v^2_b(x_s=0)$ and $v^2_b(x_s=1)$ by simulating a
particle of radius $R_i$ in a sea of particles of radius $R_j$.  Using
Eq.~(\ref{eq:thermo1}), we can calculate for these limiting values of
$x_s$ the four parameters $P(m_i,m_j)$.  We have verified that
$P(m_i,m_i)$ is independent of the type of particle. We have found
that $P(m_s,m_s)= P(m_b,m_b)= 0.145$, $P(m_b,m_s)=0.229$ (at $x_s=1$)
and $P(m_s,m_b)=0.066$ (at $x_s=0$).

To a first approximation, we consider the $P(m_i,m_j)$ to be
independent of the relative velocity. We compare in figure
\ref{fig:vsvb-xs} the values of $v^2_s$ and $v^2_b$ obtained from the
numerical simulations and those deduced from
Eq.~(\ref{eq:thermo1}). These values were calculated for different
$x_s$ at a packing fraction of $40 \%$. The dashed lines correspond to
the theoretical values and the symbols to the simulations. The
agreement between simulations and theory is quite good, in particular
for the big particles.

However, the energy lost in a collision does depend on the relative
velocity, and therefore on $v_s^2$ and $v_b^2$. To treat this
correctly in Eq.~(\ref{eq:thermo1}), we show in figure
\ref{fig:E-xs_more} that both, the kinetic energy of the system and
${v^2_s}/{v^2_b}$, decrease linearly with $x_s$. Recalling that $x_s$
lies in the interval [0,1], we can thus write:
\begin{equation}
\label{eq:thermo3}
\left\lbrace
\begin{array}{l}
\frac{C x_s}{\pi R_s^2}m_s v^2_s(x_s) + \frac{C (1-x_s)}{\pi
R_b^2}m_b v^2_b (x_s) = \frac{C}{\pi R_b^2} m_b v^2_b(x_s=0) + \left[
\frac{C}{\pi R_s^2} m_s v^2_s(x_s=1) -   \frac{C}{\pi R_b^2} m_b
v^2_b(x_s=0)\right]x_s, \\
\frac{v^2_s(x_s)}{v^2_b(x_s)} =
\frac{v^2_s(x_s=0)}{v^2_b(x_s=0 )} +
\left[\frac{v^2_s(x_s=1)}{v^2_b(x_s=1 )}
-\frac{v^2_s(x_s=0)}{v^2_b(x_s=0 )}\right]x_s.
\end{array}
\right.
\end{equation}
Using the four equations \ref{eq:thermo3} and \ref{eq:thermo1}, we can
calculate directly $v^2_s$, $v^2_b$, $P(m_s,m_b)$ and $P(m_b,m_s)$ as
a function of $x_s$. Note that different values of $x_s$ correspond to
different values of the ratio ${v^2_s}/{v^2_b}$. The solid lines in figure
\ref{fig:vsvb-xs} correspond to the theoretical
velocities squared calculated with this approach.  The agreement with
simulations is now very good.  The values of $P(m_s,m_b)$ and
$P(m_b,m_s)$ from the solution of our four equations are shown in
figure \ref{fig:PP}. The values of $P(m_s,m_b)$ for $x_s$ near 1 and
$P(m_b,m_s)$ for $x_s$ near $0$ should be taken with precaution,
because their weight in the balance of energy (Eq.~(\ref{eq:thermo1}))
is negligeable for $x_s
\simeq 1$ and $x_s \simeq 0$. Indeed the energy loss in these limit
cases corresponds to rare collisions.

Note that the approximation of $P(m_b,m_s)$ by a constant (as done in
the first approach) is fairly good. However, the value of $P(m_s,m_b)$
varies significantly with $x_s$. This explains the slight difference
between simulation and theory in the first approach.  Finally we see
that at large $x_s$ (corresponding to small ratio
${v^2_s}/{v^2_b}$) the small particles gain energy in collisions with
big particles. This phenomenon occurs only in dissipative forced gases
where $v^2$ is no longer proportional to $1/m$.

\subsection{Non-uniform agitation}
\label{sec:bidi-S}

We will now treat the case of  non-uniform agitation. To do this
we use exactly the same algorithm but introduce a gradient in the
agitation by imposing the following spatial dependence for $\eta_0^2$:
\begin{equation}
\label{eq:grad}
\eta_0^2(x) = a + b x,
\end{equation} 
where $a$ and $b$ are constants. The initial configuration of the
system is taken to be the stationary state found in the case of
homogeneous agitation. The gradient of the agitation is chosen such
that the mean value of $\eta_0^2(x)$ over the whole system corresponds
to the agitation of the initial state.  The simulations show that a
concentration gradient appears in the system. The system reaches a
non-uniform steady state where the density gradient remains present in
the course of time.  Note, however, that if we follow the motion of a
particle, we find that it does visit the whole system. All the
previous relations and equations are still valid in this case but one
should consider a local ``equilibrium''.  Figure \ref{fig:segre}
represents a typical configuration obtained at large time. The hot
agitation is on the right side of the system. In the stationary state,
the gradient of concentration balances the gradient of agitation such
that the pressure is homogeneous throughout the system.

The other main observation is that segregation appears in the system.
The local proportion of $s$ and $b$ grains is no longer the initial
one. The big particles are more sensitive to the gradient of agitation
than the small ones.  We show in figure \ref{fig:grad} the local
packing fraction for both species as a function of the position $x$ in
the system. We have found in all cases we have investigated so far
that the stationary state always exhibits segregation. Our numerical
results on segregation (the big particles are more concentrated in the
colder region) is in complete agreement with the theoretical
calculations based on the granular kinetic theory \cite{boys}.

\section{Conclusion}
\label{sec:conclu}
The main purpose of this paper is to investigate granular mixtures
numerically. We used an algorithm which keeps the dissipative
particles agitated by applying to the grains an external random
acceleration independent of their mass.  In the case where the
external acceleration is independent of the position of the particles
(homogeneous agitation) we have shown that the system reaches a well
mixed homogeneous stationary state even in the bidisperse case. We
have therefore shown that agitation, and therefore diffusion, is an
efficient mechanism for mixing.  We have established a theoretical
expression between the typical time of mixing, $\tau_{mix}$, and the
number $N_{s,b}$ of collisions between $s$ and $b$ grains per unit of
time which is valid for monodisperse as well as bidisperse
systems. For the monodiperse case we have given the exact expression
of $\tau_{mix}$ as a function of the diffusion coefficient.  For the
bisperse case we have found that $\tau_{mix}$ depends strongly on the
initial configuration of the system.  We have characterized the steady
state reached by bidisperse assemblies and in particular we have
established energetic relations which allow us to evaluate the square
velocities of the grains as a function $x_s$ (the relative proportion
of both species).

In addition we have investigated the case where there is a gradient of
agitation through the system and have shown that segregation
appears. The main cause of this segregation is not related directly to
the nature of the grains (size, dissipation, roughness) but originates
from the presence of a temperature gradient.  In a lot of mixing
experiments (rotating drum, vibrated system) there often exists a
gradient of agitation which then leads to a segregation process. Of
course there are other types of segregation mechanisms which are
sometimes more efficient, but this one is in some sense
intrinsic.\vspace{3ex}

\centerline{\bf Acknowledgments}
\vspace{3ex}

This work was partially funded by the CNRS Programme International de
Cooperation Scientifique PICS $\#753$. C. H. thanks Alexandre Valance
for his support during this work and his help for writing of this
paper.



\begin{figure}
\caption{Configurations at three different times $t=1$, $t=50$ and
$t=199$. {\bf a)}: for an initially mixed system and {\bf b)}: for an
initially segregated system. }
\label{fig:mono}
\end{figure}

\begin{figure}
\caption{Number of collisions per unit time between the two species as
a function of time. The dashed (solid) line corresponds to the case of
FIG.~\ref{fig:mono}{\bf a} ({\bf b}).}
\label{fig:N-t_mono}
\end{figure}

\begin{figure}
\caption{$N_{s,b}(\infty) - N_{s,b}(t)$ vs
time. Solid line is the numerical average over 5 simulations. The dashed line is given by 
$N_{s,b}(\infty)exp(-t/\tau_{mix})$, $\tau_{mix}= 43.5$.}
\label{fig:n-fit}
\end{figure}

\begin{figure}
\caption{Dependence of $\tau_{mix}$ on $D$. $\bigcirc$: value from
simulations. Solid line: theoretical prediction. }
\label{fig:dep-D}
\end{figure}

\begin{figure}
\caption{Evolution, at short time, of an initially segregated
bidisperse system; From left to right $t=1$, $t=6$ and $t=36$. $R_s=
0.4$ and; $R_b=0.6$, the total packing fraction is $40\%$. $x_s=0.5$.}
\label{fig:etero}
\end{figure}

\begin{figure}
\caption{The local packing fraction of the small particles as
a function of $x$ (the lateral position) and $t$.}
\label{fig:vague}
\end{figure}

\begin{figure}
\caption{ $v^2_s(t)$ and $v^2_b(t)$ as functions of time,
for an initially segregated system.}
\label{fig:vit-t}
\end{figure}

\begin{figure}
\caption{$N_{s,b}(t)$ vs $t$ for a bidisperse gas. $R_s=0.4$, 
$R_b=0.6$, and $x_s=0.625$. $\tau_{mix}
\simeq 30$ is obtained by fitting Eq.~(\ref{eq:N-exp}), shown as a
dashed line.}
\label{fig:N-t_etero}
\end{figure}

\begin{figure}
\caption{$\tau_{mix}$ for the bidisperse case as a function
of $x_s$. $R_s=0.4$,$R_b=0.6$ and $C= 0.4$. $\tau_{mix}$ is constant.}
\label{fig:t_mixt_etero}
\end{figure}

\begin{figure}
\caption{The mean square velocities reached in the stationary state as
a function of $x_s$.  $v_s^2$ ($\Box$) and $v_b^2$ ($\bigcirc$)
obtained from simulations. Dashed lines are obtained from
eq~\ref{eq:thermo1} assuming constant $P(m_i,m_j)$'s. Solid lines are
the results from Eq.~(\ref{eq:thermo1}) and Eq.~(\ref{eq:thermo3}).}
\label{fig:vsvb-xs}
\end{figure}

\begin{figure}
\caption{ a) The total kinetic energy of the system as a function of
$x_s$ for $R_s=0.4$, $R_b=0.6$, $C=0.4$ and $\eta_0^2=22.0$.  b)
${v^2_s}/{v^2_b}$ as a function of $x_s$ for the same parameters.}
\label{fig:E-xs_more}
\end{figure}

\begin{figure}
\caption{Values of $P(m_s,m_b)$ and $P(m_b,m_s)$ vs $x_s$ calculated 
using eqs \ref{eq:thermo3} and \ref{eq:thermo1} for the same mixtures
as in FIG.~\ref{fig:E-xs_more} Dashed line: $P(m_b,m_s)$. Full line:
$P(m_s,m_b)$. }
\label{fig:PP}
\end{figure}

\begin{figure}
\caption{Typical configuration obtained in the presence of a gradient
in the granular temperature. $R_s=0.4$, $R_b=0.6$, $x_s= 0.5$, $a=4$
and $b=0.9$. $C=0.3$.}
\label{fig:segre}
\end{figure}

\begin{figure}
\caption{Local packing fraction as a function of $x$ for $s$ ($\Box$) 
and $b$ ($\bigcirc$) grains. The mixture corresponds to
FIG.~\ref{fig:segre}. Averaged over the whole system, $x_s=0.5$ and
$C=0.3$.}
\label{fig:grad}
\end{figure}
\end{document}